\documentclass{INTERSPEECH2023}
\usepackage{amsmath,graphicx}
\usepackage{cite,subfigure,amssymb,tablefootnote,multirow,tipa}
\pdfoutput=1

\interspeechcameraready 

\title{OTF: Optimal Transport based Fusion of Supervised and Self-Supervised Learning Models for Automatic Speech Recognition}
\name{Li Fu, Siqi Li, Qingtao Li, Fangzhu Li, Liping Deng, Lu Fan, \\ Meng Chen, Youzheng Wu, Xiaodong He}
\address{JD AI Research, Beijing, China}
\email{\{fuli3,lisiqi26,liqingtao8,lifangzhu1,dengliping6,fanlu\}@jd.com
\{chenmeng20,wuyouzheng1,hexiaodong\}@jd.com}

\begin{document}
\maketitle

\begin{abstract}
Self-Supervised Learning (SSL) Automatic Speech Recognition (ASR) models have shown great promise over Supervised Learning (SL) ones in low-resource settings. However, the advantages of SSL are gradually weakened when the amount of labeled data increases in many industrial applications. To further improve the ASR performance when abundant labels are available, we first explore the potential of combining SL and SSL ASR models via analyzing their complementarity in recognition accuracy and optimization property. Then, we propose a novel \underline{O}ptimal \underline{T}ransport based \underline{F}usion (OTF) method for SL and SSL models without incurring extra computation cost in inference. Specifically, optimal transport is adopted to softly align the layer-wise weights to unify the two different networks into a single one. Experimental results on the public 1k-hour English LibriSpeech dataset and our in-house 2.6k-hour Chinese dataset show that OTF largely outperforms the individual models with lower error rates.
\end{abstract}

\noindent\textbf{Index Terms}: Automatic speech recognition, model fusion, optimal transport, self-supervised learning

\section{Introduction}
\label{sec:introduction}
Recently, Self-Supervised Learning (SSL) has emerged as a successful paradigm to address the issue of label scarcity in low-resource Automatic Speech Recognition (ASR) tasks, e.g. multiple languages~\cite{babu2021xls,khurana2022magic,zhao2022improving} and domain shift~\cite{zuluaga2023does,thomas2022efficient}. It usually pre-trains a representation model on numerous unlabeled utterances, and then fine-tunes the model with a relatively small amount of labeled speech~\cite{baevski2020wav2vec,hsu2021hubert,chen2022wavlm,mohamed2022self,baevski2020effectiveness,yang2021superb,chung2021w2v}. Nevertheless, the gains achieved by the pre-training might diminish when the amount of downstream labeled dataset increases~\cite{zhang2022bigssl,wang2021unispeech,radford2022robust}. Thus, there would be a dilemma in many resource-rich industry applications -- Shall we train a Supervised Learning (SL) ASR model from scratch or fine-tune the pre-trained representation model?

In general, SL models are optimized to perform well when large amounts of labeled and in-domain speech are available~\cite{li2020popular,gulati2020conformer,fu2021scala}; and SSL models are considered to have good generalization via pre-training on numerous unlabeled utterances~\cite{hsu2021robust,zan2022complementarity,bucci2021self}. While the performance gaps in Word/Character Error Rate (WER/CER) between the SL and SSL ASR models become small in resource-rich settings, we believe that empirically, the two models would contain diverse or even complementary abilities due to the training process being quite different. Based on this idea, as analyzed in Sec.~\ref{subsec:analysis}, we first verify the potential of fusing SL and SSL models according to 1) Recognition accuracy: We roughly estimate the {\it upper bound} of model fusion via picking out the best hypothesis (HYP) of different models for each test sample, which implies a large improvement room on fusing these models; and 2) Optimization property: We study the optimization process of the two models via loss landscape~\cite{li2018visualizing}, which qualitatively indicates the SSL and SL models' advantages in generalization and in-domain task learning, respectively.

\begin{figure}[t]
  \centering
  \includegraphics[width=0.475\textwidth]{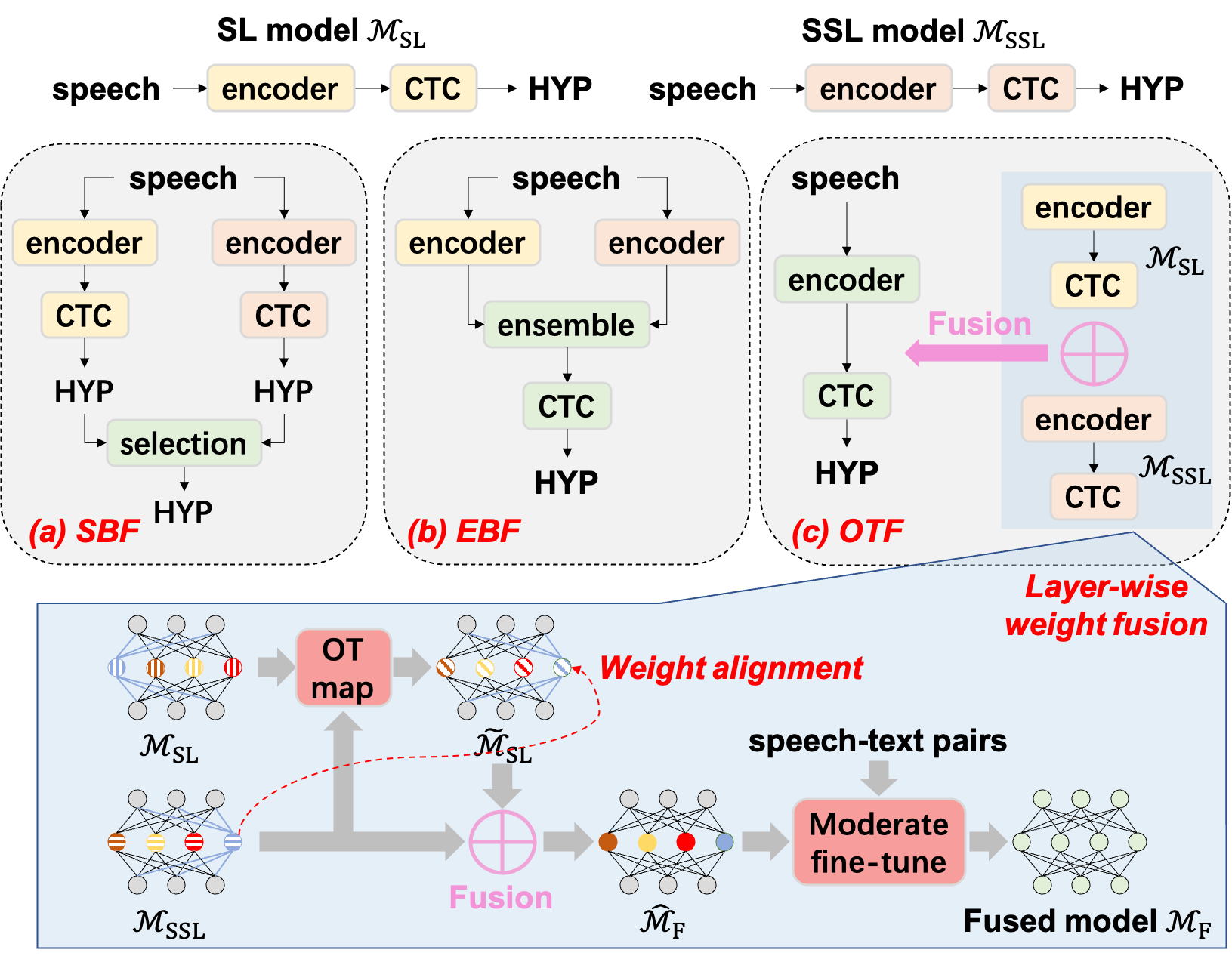}
  \caption{Overview of the proposed (c) OTF over the existing methods (a) SBF and (b) EBF: each layer's weights of SL model $\mathcal{M}_{SL}$ are aligned with SSL model $\mathcal{M}_{SSL}$ via Optimal Transport (OT) maps to yield aligned model $\widetilde{\mathcal{M}}_{SL}$, which is added with $\mathcal{M}_{SSL}$ and then moderately fine-tuned on labeled speech to obtain the fused model $\mathcal{M}_{F}$ with the same architecture.}
  \label{fig1}
\end{figure}

Existing model fusion methods mainly focus on how to aggregate the constituent models' output hypotheses or latent features. For example, Selection-Based Fusion (SBF) method (see Fig.~\ref{fig1}(a)) was proposed to select the best hypothesis from different models via the designed criteria, e.g. confidence score~\cite{soto2016selection}. Ensemble-Based Fusion (EBF) method (see Fig.~\ref{fig1}(b)) was explored to combine the latent features of multiple encoders pre-trained in different frameworks~\cite{arunkumar2022investigation,tang2022exploring,wu2022efficacy,sun2022stacking}. However, both the SBF and EBF methods would inevitably suffer high computational cost at test time since each utterance is processed through all of these encoders. A naive solution might be aggregating all the models into a single one by directly averaging the model parameters. However, since the weight parameters of SL and SSL models are not one-to-one corresponded, such direct averaging is ineffective and may even damage the well-trained models. To circumvent this issue, matching weight before averaging was investigated in the computer vision domain~\cite{yurochkin2019bayesian,singh2020model}. For example, Yurochkin et al.~\cite{yurochkin2019bayesian} proposed the Bayesian nonparametric matching method to align and average the image classification models on different edge devices for federated learning. Singh et al.~\cite{singh2020model} explored the fusion of image classification models by leveraging optimal transport to align each layer's weight. However, discussion about the weight-based fusion for ASR models is quite rare, which might be more challenging than the image classification tasks~\cite{yurochkin2019bayesian,singh2020model} since speech signals are sequential values~\cite{hsu2021hubert}.

To combine the SL and SSL models without increasing the inference cost, we propose an \underline{O}ptimal \underline{T}ransport based \underline{F}usion (OTF) method, which fuses the two ASR models into the same architecture and improves the performance in the following way (see Fig.~\ref{fig1}(c)). First, inspired by the work of~\cite{singh2020model}, we adopt optimal transport to softly align each layer's weights of the SL and SSL models. Specifically, a layer-wise transport map is estimated via minimizing the cost of transferring the distribution of the SL model's weight to the SSL model. Then, the input and output parameters of the SL model's weight are aligned by multiplying with the transport maps of preceding and current layers, respectively. Finally, to enhance the ASR performance, the aligned SL model is averaged with the SSL model and moderately fine-tuned on labeled data to obtain the fused model. Extensive experiments on different datasets show that OTF effectively fuses the SL and SSL models with obvious WER/CER reductions. Our main contributions are summarized as follows:

\begin{itemize}[leftmargin=0.12in]
	\setlength{\itemsep}{-2pt}
	\item To the best of our knowledge, this is the first work exploring the fusion of SL and SSL models for speech recognition.
       \item We propose a novel approach, named OTF, which unifies SL and SSL ASR models efficiently without incurring an extra computational cost in inference. 
	\item We verify the effectiveness of OTF, with discussion, on English and Chinese datasets with large WER/CER reductions compared with the individual models and baseline methods.
\end{itemize}


\section{Related Work}
\label{sec:related_work}

\noindent{\bf Model fusion.} In recent years, it has been shown favorable to fuse different models to enhance the ASR performance. Although the SBF method was explored to pick the better hypothesis of two ASR models without re-training, it would result in a suboptimal solution since the criteria used for hypothesis selection might not be accurate for each utterance~\cite{soto2016selection,qiu2021learning}. More recently, the EBF method was proposed to tightly couple two pre-trained models. Arunkumar et al.~\cite{arunkumar2022investigation} investigated an ensemble model to combine the outputs of the last layer of HuBERT~\cite{hsu2021hubert} and wav2vec2~\cite{baevski2020wav2vec} fine-tuned ASR models. Tang et al.~\cite{tang2022exploring} explored the combination of the multi-layer latent features of wav2vec2~\cite{baevski2020wav2vec} and data2vec~\cite{baevski2022data2vec}. Wu et al.~\cite{wu2022efficacy} proposed an ensemble framework, with a combination of ensemble techniques to fuse SSL speech models’ embedding. However, both the SBF and EBF methods would suffer high computational cost at test time with aggregating all the models’ parameters. In contrast, our proposed OTF fuses the two ASR models into the same architecture via weight alignment, which will not incur an extra cost during inference.

\noindent{\bf Multi-task learning.} Another way to incorporate auxiliary abilities to the ASR model is multi-task learning. For example, distillation tasks~\cite{hinton2015distilling} were added with ASR tasks to transfer the teacher model's (e.g. SSL model's) knowledge to the student model (e.g. SL model)~\cite{huang2023ensemble,cao2021improving,takashima2019investigation}. The existing methods usually assume the teacher model is superior to the student model, while the performance gaps of the SL and SSL models might be small in resource-rich settings. The joint of SL and SSL training losses for ASR tasks were studied in~\cite{talnikar2021joint,wang2021unispeech_icml,bai2022joint}. However, as mentioned in~\cite{bai2022joint}, it might be difficult to balance the SSL and SL components systematically. Differently, our work focuses on weight-based fusion, which is more efficient to inherit the abilities of the individual models. Moreover, the multi-task learning method might also complement the moderate fine-tuning in the proposed OTF, which will be investigated in our future work.

\section{Our Proposed Approach}
\label{sec:Approach}

\begin{figure}[t]
	\centering
	\subfigure[English ASR models]{
		\begin{minipage}[t]{0.46\linewidth}
			\centering
			\includegraphics[width=\linewidth]{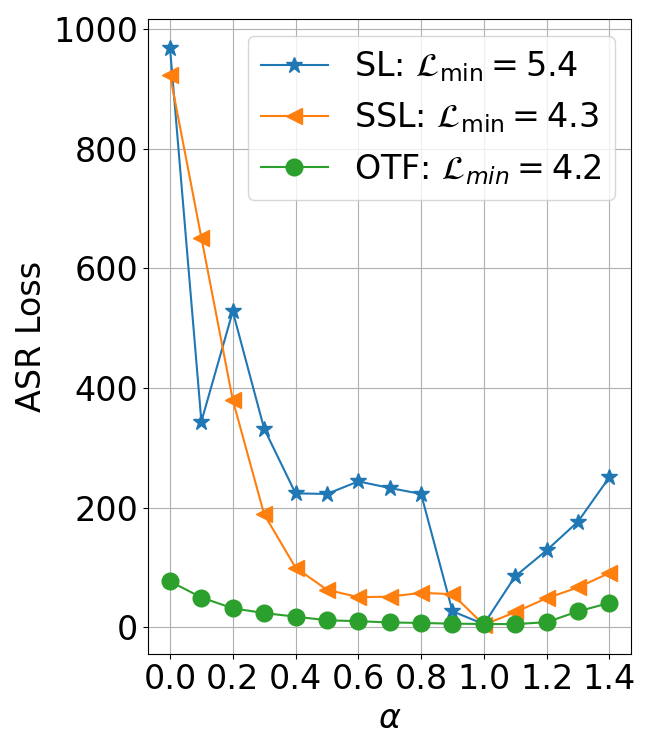}
		\end{minipage}
	}%
	\subfigure[Chinese ASR models]{
		\begin{minipage}[t]{0.46\linewidth}
			\centering
			\includegraphics[width=\linewidth]{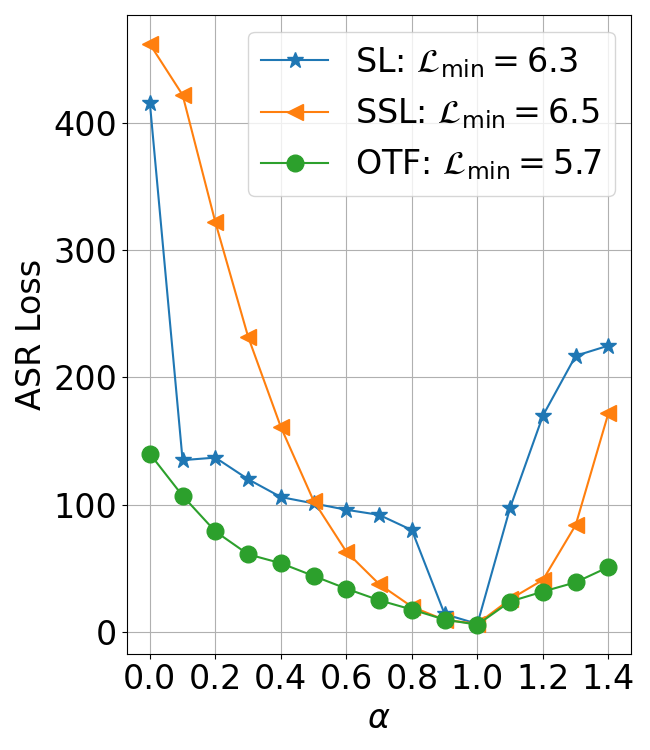}
		\end{minipage}
	}%
	\centering
	\caption{Loss landscapes of the linear interpolation along the three kind of models ($\alpha=1$): 1) SL models, 2) SSL models, 3) fused models via our proposed OTF, with their initial models ($\alpha=0$), respectively, where $\alpha$ is the interpolation coefficient.}
	\label{fig3}
\end{figure}

\subsection{Advantages for fusing SL and SSL ASR models}

\label{subsec:analysis}
\noindent{\bf Complementarity in recognition accuracy.} Although the gaps in WER/CER of the SL and SSL ASR models\footnote{Details about the models are shown in Sec.~\ref{subsec:exp_setup}.} are small in resource-rich settings, we find that the recognition errors of the two models might be different for a certain utterance.
The potential gain by the fusion of SL and SSL models can be estimated via picking the best hypothesis among the individual models for each test utterance to calculate the error rate performance.
It can be regarded as an approximated upper bound of model fusion, as shown in Table~\ref{table1}-\ref{table2}.
Numerically, compared with the best among SL and SSL models, the relative WER/CER reduction of the upper bound can lead up to 21$\%$ for the English and Chinese models. This implies a large potential for improvement for the model fusion as we expected.

\noindent{\bf Complementarity in optimization property.} To find out the advantages of SL and SSL models with abundant speech labels, we analyze the optimization properties of the two models via the visualization of loss landscape~\cite{li2018visualizing}. Specifically, given the weights of an SL/SSL ASR model $\theta$ and its initialized weights $\theta_{0}$ before fine-tuning, we plot the loss landscape $\mathcal{L}(\theta(\alpha))$ of the validation dataset, with $\alpha$ the coefficient for the weights' linear interpolation $\theta(\alpha)=(1-\alpha)\theta_{0}+\alpha\theta$; and with $\mathcal{L}$ the ASR loss function used for fine-tuning. As shown in Fig.~\ref{fig3}, the curves of the loss landscape show that each of SL and SSL has its advantage in the optimization process: 1) The flat minima points of SSL models present a better generalization than the SL models' sharp minima points~\cite{keskarlarge2017minima}; and 2) The Chinese SL model achieves a better minima (and lower CERs on Chinese Cantonese test sets in Table~\ref{table2}) than the SSL model when large amounts of in-domain labeled speech is available. The results qualitatively indicate the SSL and SL models’ complementarity in generalization and in-domain task learning, respectively.

\begin{figure}[t]
    \vspace{-0.2in}
  \centering
  \includegraphics[width=0.465\textwidth]{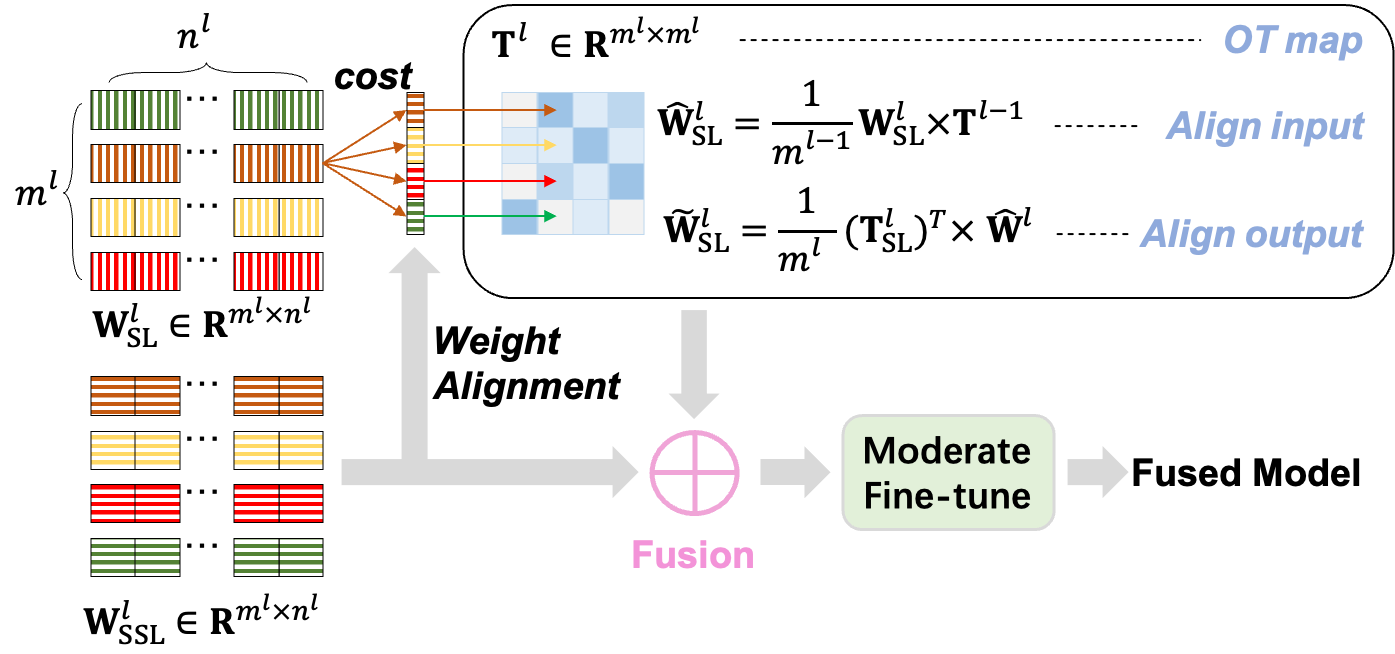}
  \caption{Illustration of the layer-wise weight alignment and fusion via optimal transport. The $l$-th layer's weight matrix $\mathbf{W}_{\rm SL}^{l}$ of the SL model is aligned with weight matrix $\mathbf{W}_{\rm SSL}^{l}$ of the SSL model with the transport maps.}
  \vspace{-0.1in}
  \label{fig4}
\end{figure}

\subsection{Model fusion by optimal transport}
\noindent{\bf Key intuition:} Our work aims to fuse the SL and SSL ASR models into a single model with the same architecture while taking advantage of both models. To achieve this, we propose OTF, which matches similar weights in each layer for model aggregation (see Fig~\ref{fig4}). Instead of searching over the space of permutation matrices, transport mapping matrices are calculated to softly align one model's weight to another model according to the cost (e.g. Euclidean distance~\cite{wang2020federated}) across the two weights' row vectors~\cite{singh2020model}. Then each layer's input and output are aligned by the maps of the preceding and current layers, respectively.

Given an SL model $\mathcal{M}_{\rm SL}$ and an SSL model $\mathcal{M}_{\rm SSL}$ that are composed of $L$ layers of weight $\mathbf{W}_{\rm SL}^{l}\in{\mathbf{R}^{m^l\times{n^l}}}$ and $\mathbf{W}_{\rm SSL}^{l}\in{\mathbf{R}^{m^l\times{n^l}}}$ with $l\in[1,2,\cdots,L]$, $n^l$ and $m^l$ the dimensions\footnote{Following~\cite{singh2020model}, the weights of convolution and bias layers are reshaped to the 2-dimensional regular weights.} of input and output, respectively. Note that the input dimension of the current layer is equal to the output dimension of the preceding layer. Then the transport map of the $l$-th layer $\mathbf{T}^{l}\in\mathbf{R}^{m^l\times{m^l}}$ is estimated by minimizing the cost $\mathcal{L_{\rm t}}$ in transferring the weight distribution of $\mathbf{W}_{\rm SL}^{l}$ to $\mathbf{W}_{\rm SSL}^{l}$~\cite{singh2020model},

\vspace{-0.05in}
\begin{equation}
    \mathcal{L_{\rm t}}=\underset{\mathbf{T}^l}{\min}  \left \langle \mathbf{T}^l,\mathbf{D}^l \right \rangle _{F}
    \label{eq1}
    \vspace{-0.05in}
\end{equation}

\noindent{where $\left \langle \cdot,\cdot \right \rangle _{F} $} is the Frobenius inner product, $\mathbf{D}^l\in\mathbf{R}^{m^l\times{m^l}}$ is the Euclidean distance matrix of the rows in $\mathbf{W}_{\rm SL}^{l}$ to $\mathbf{W}_{\rm SSL}^{l}$. $\mathbf{T}^l\mathbf{1}_{m^l}$ and $({\mathbf{T}^l})^{T}{\mathbf{1}_{m^l}}$ are constrained to be equal to the $m^l$-dimensional uniform distributions. The optimization is to find a better way (i.e. $\mathbf{T}^{l}$) to move the $l^{th}$ layer of the SL model's weight to the SSL model's by minimizing the defined cost -- If a row of the SL model's weight matrix is similar to a row of the SSL model's weight, the transportation cost is defined as small. The Frobenius inner product is the sum of the cost to move the SL model's weight to the SSL model's. Since our method fuses the two models' weight directly, we do not need any data to perform the optimization. Note that the data would be alternatively used to align the output features of each layer~\cite{singh2020model}. 

Denoting the map of the preceding layer as $\mathbf{T}^{l-1}$, the input weights are aligned to obtain the intermediate matrix as~\cite{singh2020model}:
\vspace{-0.05in}
\begin{equation}
    {\widehat{\mathbf{W}}_{\rm SL}^{l}=\frac{1}{m^{l-1}} {\mathbf{W}}_{\rm SL}^{l}\times {\mathbf{T}}^{l-1}}
    \label{eq2}
    \vspace{-0.05in}
\end{equation}

Then, the output weights are aligned via pre-multiplying $\widehat{\mathbf{W}}_{\rm SL}^{l}$ by the transposition of $\mathbf{T}^{l}$, yields
\vspace{-0.075in}
\begin{equation}
    {\widetilde{\mathbf{W}}_{\rm SL}^{l}=\frac{1}{m^{l}} {(\mathbf{T}}^{l})^T}\times{\widehat{\mathbf{W}}_{\rm SL}^{l}}
    \label{eq3}
\end{equation}

Finally, we calculate the fused weight of the $l$-th layer by $\widehat{\mathbf{W}}_{\rm F}^{l}=\frac{1}{2}(\mathbf{W}_{\rm SLL}^{l}+{\widetilde{\mathbf{W}}_{\rm SL}^{l})}$, which is fine-tuned on the labeled dataset to improve the recognition performance. Since the fused model $\mathcal{M}_{\rm F}$ is averaged with the weight alignment, it can still partly recognize the speech and only moderate/limited fine-tuning steps are needed in this stage. The effectiveness of weight alignment and fine-tuning are studied in Sec.~\ref{subsec:discussion}.

\begin{table}[t]
        \vspace{-0.2in}
	\centering  
	\caption{WER performance  of English systems and relative WER improvement of OTF over SSL model (in brackets) ($\%$).}
	\label{table1}  
	\resizebox{1.00\linewidth}{!}{
	\begin{tabular}{l|c|c|c|c|c|c|c}  
		\hline  
            \multirow{2}{*}{Methods} &model& \multicolumn{2}{c|}{training data (h)} & \multicolumn{2}{c|}{test-clean} & \multicolumn{2}{c}{test-other} \\
            \cline{3-8}
            ~&size&unlabeled&labeled&\multicolumn{1}{c|}{offline} & \multicolumn{1}{c|}{online} & \multicolumn{1}{c|}{offline} & \multicolumn{1}{c}{online} \\
		\hline
		\hline
		  SL~\cite{yao2021wenet} & 0.1B&NA & 1k & 3.6& 4.7&9.7&12.4  \\
            SSL~\cite{fu2022ufo2} & 0.1B& 1k & 1k &3.3&4.2&7.7&10.3  \\
            \hline
            Upper- & \multirow{2}{*}{0.2B}& \multirow{2}{*}{NA} & \multirow{2}{*}{NA} &\multirow{2}{*}{\bf 2.6}&\multirow{2}{*}{\bf 3.4}&\multirow{2}{*}{\bf 6.7}&\multirow{2}{*}{\bf 9.0}\\
            bound & &  &  &&&&\\
            \hline
            \hline
            SBF~\cite{soto2016selection} & 0.2B& NA & NA & 3.3&4.1&7.9&10.3  \\
            EBF~\cite{arunkumar2022investigation} & 0.2B& NA & 1k & 3.3&	4.3&7.7&10.2 \\
            OTF & 0.1B& NA & 1k & {\bf 3.0(+9.1)}  & {\bf 4.0(+4.8)} & {\bf 7.4(+3.9)}  & {\bf 10.1(+1.9)}  \\
		\hline
	\end{tabular}
	}
\end{table}

\begin{table}[t]
	\centering  
	\caption{CER performance  of Chinese systems and relative CER improvement of OTF over SSL model (in brackets) ($\%$).}
	\label{table2}  
	\resizebox{1.00\linewidth}{!}{
	\begin{tabular}{l|c|c|c|c|c|c|c}  
		\hline  
            \multirow{2}{*}{Methods} &model& \multicolumn{2}{c|}{training data (h)} & \multicolumn{2}{c|}{test-Mandarin} & \multicolumn{2}{c}{test-Cantonese} \\
            \cline{3-8}
            ~&size&unlabeled&labeled&\multicolumn{1}{c|}{offline} & \multicolumn{1}{c|}{online} & \multicolumn{1}{c|}{offline} & \multicolumn{1}{c}{online} \\
		\hline
		\hline
		  SL~\cite{yao2021wenet} & 0.1B&NA & 2.6k & 12.8 &15.7&12.6&13.8  \\
            SSL~\cite{fu2022ufo2} & 0.1B& 150k & 2.6k &12.2	&14.5&12.9&14.0 \\
            \hline
            Upper- & \multirow{2}{*}{0.2B}& \multirow{2}{*}{NA} & \multirow{2}{*}{NA} &\multirow{2}{*}{\bf 10.1}&\multirow{2}{*}{\bf 11.9}&\multirow{2}{*}{\bf 10.6}&\multirow{2}{*}{\bf 12.1}\\
            bound & &  &  &&&&\\
            \hline
            \hline
            SBF~\cite{soto2016selection} & 0.2B& NA & NA & 11.6&13.5&12.1&13.8  \\
            EBF~\cite{arunkumar2022investigation} & 0.2B& NA & 2.6k & 11.6&	13.5&12.2&13.8 \\
            OTF & 0.1B& NA & 2.6k & {\bf 11.1(+9.0)}  & {\bf 12.7(+12.4)} & {\bf 11.8(+8.5)}  & {\bf 13.1(+6.4)}  \\
		\hline
	\end{tabular}
	}
        \vspace{-0.15in}
\end{table}

\section{Experiments and Discussion}
\label{sec:experiments}
\subsection{Experimental setup}
\label{subsec:exp_setup}
{\bf Data preparation.} Two datasets in different languages are used in our experiments: 1) English data: the public LibriSpeech~\cite{panayotov2015librispeech} which contains 1k-hour speech-text pairs; and 2) Chinese data: our in-house 150k-hour Mandarin-Cantonese dataset and only 2.6k-hour data are labeled, with a ratio of Mandarin to Cantonese 10:1. For LibriSpeech, the WER performance of the ASR model is evaluated on the original test-clean (considered easier) and test-other (considered harder and noisier) datasets. For the other one, 5$\%$ of the labeled Mandarin and Cantonese samples are randomly selected as the test sets, named test-Mandarin and test-Cantonese. The 80-dimensional Mel-spectrograms of each utterance are pre-processed as the input to the model with 25ms window size and 10ms step size.\\
{\bf Models.} Considering the preference of unified offline and online ASR in industrial applications~\cite{yao2021wenet,yu2020dual,fu2022ufo2}, we use the State-Of-The-Art (SOTA) SSL-based unified framework -- UFO2~\cite{fu2022ufo2} to test the performance of OTF in both offline and online modes. The model consists of 2 convolutional sub-sampling layers and 12 Conformer blocks with dimension 512 for the encoder. The number of model parameters is 0.1 billion. More details about the model architecture can be referred to~\cite{fu2022ufo2}. Here, we focus on the fusion of SL and SSL models in two scenarios as below:

\noindent{\it 1) English ASR models}
\begin{itemize}[leftmargin=0.2in]
	\setlength{\itemsep}{-2pt}
	\item $\mathcal{M}_{\rm SL}$: trained on the 1k-hour public LibriSpeech dataset.
        \item $\mathcal{M}_{\rm SSL}$: pre-trained and fine-tuned on the 1k-hour LibriSpeech dataset successively.
\end{itemize}

\noindent{\it 2) Chinese ASR models}
\begin{itemize}[leftmargin=0.2in]
	\setlength{\itemsep}{-2pt}
	\item $\mathcal{M}_{\rm SL}$: trained on the 2.6k-hour labeled Chinese dataset.
        \item $\mathcal{M}_{\rm SSL}$: pre-trained on the 150k-hour unlabeled Chinese dataset and fine-tuned on the 2.6k-hour dataset successively.
\end{itemize}

Both the pre-training and fine-tuning are optimized with a mini-batch of 96 and using the same learning rate scheduler as~\cite{fu2022ufo2}. Note that the SL and SSL models are fully fine-tuned with 200 epochs and 80 epochs for the English and Chinese tasks, respectively. The epoch number in the moderate fine-tuning of the proposed OTF is set to 10 for the two languages.

\noindent{\bf Decoding.} Following the representative SSL works~\cite{baevski2020wav2vec,hsu2021hubert,baevski2022data2vec}, we evaluate the performance of all models using the Connectionist Temporal Classification (CTC) beam search decoder~\cite{graves2006connectionist} with 10 the decoding beam size. Note that no language model is applied in our experiments. 

\vspace{-0.05in}
\subsection{Main experiment}
\vspace{-0.05in}
\label{subsec:main_exp}
\noindent{\bf Baseline methods.} To evaluate the performance of our method on the fusion of SL and SSL models, we implement the SOTA EBF method via combining the outputs of the two models' encoders~\cite{arunkumar2022investigation}, and the existing SBF method based on the average word confidence~\cite{soto2016selection} as our baseline methods.

\noindent{\bf System performance.} As shown in Table~\ref{table1}-\ref{table2}, we compare OTF with the performance of the SL and SSL models, and the baseline methods in offline and online modes. Compared with the SL and SSL models, OTF achieves a large improvement with lower WERs/CERs. Numerically, compared with the better of SL and SSL models, OTF yields relative WERs/CERs reductions up to $9.1\%$ and $12.4\%$ in English and Chinese, respectively. Although SBF and EBF are effective in model fusion, they suffer from a high inference cost with aggregating the two models' parameters. Besides, we find the baseline methods can cause a worse performance in some test sets. We infer 1) the confidence scoring of SBF might be not accurate enough for hypothesis selections; and 2) it would be challenging for EBF to align the features of two adequately-trained but diverse models. Compared with the baselines, our method consistently achieves the best performance in both the English and Chinese scenarios. Moreover, since each layer of the individual models' weights are fused into a single one, OTF is more efficient without increasing the model size than the baseline methods~\cite{soto2016selection,arunkumar2022investigation}.

\begin{table}[t]
        \vspace{-0.2in}
	\centering  
	\caption{Ablation studies on weight alignment and Moderate Fine-Tuning (MFT) for English models in WER ($\%$).}
	\label{table3}  
	\resizebox{0.95\linewidth}{!}{
	\begin{tabular}{l|c|c|c|c|c}  
		\hline  
            \multirow{2}{*}{Methods} & \multicolumn{1}{c|}{FT} & \multicolumn{2}{c|}{test clean} & \multicolumn{2}{c}{test other} \\
            \cline{3-6}
            ~& data (h)&\multicolumn{1}{c|}{offline} & \multicolumn{1}{c|}{online} & \multicolumn{1}{c|}{offline} & \multicolumn{1}{c}{online} \\
		\hline
		\hline
    	SL & NA &3.6&4.7&9.7&12.4   \\
            \ \ \ \ \ + MFT  & 1k &3.6&4.7&9.8&12.5  \\
            		\hline
  		SSL & NA &3.3&4.2&7.7&10.3   \\
            \ \ \ \ \ + MFT  & 1k &3.3&4.3&7.7&10.3  \\
            		\hline
		  Direct Avg.  & NA & 100 &100&100&100  \\
            \ \ \ \ \ + {\it fully} FT  & 1k &3.5	&4.6&9.6&12.1 \\
            		\hline
            Aligned Avg. &NA&52.7&63.0& 78.4& 82.9\\
            \ \ \ \ \ + MFT (Our OTF) & 1k & {\bf 3.0}  & {\bf 4.0} & {\bf 7.4}  & {\bf 10.1}  \\
		\hline
	\end{tabular}
	}
        \vspace{-0.05in}
\end{table}

\begin{table}[t]
	\centering  
		\caption{Ablation studies on weight alignment and Moderate Fine-Tuning (MFT) for Chinese models in CER ($\%$).}
	\label{table4}  
	\resizebox{0.95\linewidth}{!}{
	\begin{tabular}{l|c|c|c|c|c}  
		\hline  
            \multirow{2}{*}{Methods} & \multicolumn{1}{c|}{FT} & \multicolumn{2}{c|}{test Mandarin} & \multicolumn{2}{c}{test Cantonese} \\
            \cline{3-6}
            ~&data (h)&\multicolumn{1}{c|}{offline} & \multicolumn{1}{c|}{online} & \multicolumn{1}{c|}{offline} & \multicolumn{1}{c}{online} \\
		\hline
		\hline
    	SL & NA &12.8 &15.7&12.6&13.8   \\
            \ \ \ \ \ + MFT  & 2.6k &12.7&15.6&12.8&13.9  \\
            		\hline
  		SSL & NA &12.2&14.5&12.9&14.0  \\
            \ \ \ \ \ + MFT  & 2.6k &12.1&14.4&12.9&14.1  \\
            		\hline
		  Direct Avg.  & NA & 100 &100&100&100  \\
            \ \ \ \ \ + {\it fully} FT  & 2.6k&12.9	&15.8&13.5&14.2 \\
            		\hline
            Aligned Avg. &NA& 97.5& 99.2& 92.1& 97.6\\
            \ \ \ \ \ + MFT (Our OTF) & 2.6k & {\bf 11.1}  & {\bf 12.7} & {\bf 11.8}  & {\bf 13.1}  \\
		\hline
	\end{tabular}
	}
        \vspace{-0.2in}
\end{table}

\subsection{Ablation study and discussion}
\label{subsec:discussion}
To further evaluate the effectiveness of OTF, we analyze the proposed method from the following three perspectives.

\noindent {\bf 1) Ablation study.} To analyze the effect of weight alignment and moderate fine-tuning in OTF, we compare our proposed method with: direct averaging of the SL and SSL models; SL models with moderate fine-tuning; and SSL models with moderate fine-tuning (see Table~\ref{table3}-\ref{table4}). {\bf (a) Weight alignment.} Before fine-tuning, we find that direct averaging of the two models will cause the recognition results to be all empty with 100$\%$ error rate. Although the performance can be substantially improved after fully fine-tuning on labeled data, the performance is inferior to the better one of the individual models. We infer the direct averaging of the two different models would damage the well-trained models and lead to a suboptimal solution. Differently, our method averages the layer-wise weights via alignment, which can still partly recognize the speech even without fine-tuning. We also find that the averaging after alignment for the ASR tasks achieves a larger performance degradation compared with the image classification tasks~\cite{singh2020model}. We infer it is more difficult to align the weights of sequence-to-sequence ASR models than the instance classification models. However, the proposed OTF still achieves the best performance than other methods with limited fine-tuning steps. {\bf (b) Moderate fine-tuning.} As for the moderate fine-tuning in OTF, we compared SL and SSL models with the same fine-tuning to test if the performance improvement of our method is caused by more training steps. Since the SL and SSL are adequately trained, the performance is almost the same before and after the moderate fine-tuning. The experimental results show that our method fuses the two models effectively, and can obtain a better initialization to be fine-tuned for performance enhancement.

\noindent {\bf 2) Visualization of loss landscape.} From the perspective of the optimization property, we plot the loss landscape of OTF, as shown in Fig.~\ref{fig3}. The curves of loss landscape show that our method achieves the flattest and lowest minimum points than both the SL and SSL models. It implies that OTF fuses the two models and incorporates both of the individual advantages in generalization and in-domain optimization as we expected.

\noindent {\bf 3) Performance on small labeled datasets.} As so far, we have verified the effectiveness of OTF on abundant datasets, i.e. one thousand hours and more. However, we still wonder about the performance when only a small amount of labeled dataset is available. Here, we assume that only a 100-hour train-clean subset of Librispeech~\cite{panayotov2015librispeech} and a 50-hour Cantonese dataset are labeled. As shown in Table~\ref{table5}-\ref{table6}, the relative improvement of OTF decreases when there is a large performance gap between the two models. We infer that if there is a large gap between the two constituent models, the better one will dominate the results of model fusion. Nevertheless, our method can still improve the performance of the fused models with small labeled datasets.

\begin{table}[t]
        \vspace{-0.2in}
	\centering  
		\caption{Performance on small English dataset in WER ($\%$).}
	\label{table5}  
	\resizebox{0.9\linewidth}{!}{
	\begin{tabular}{l|c|c|c|c|c|c}  
		\hline  
            \multirow{2}{*}{Methods} & \multicolumn{2}{c|}{training data (h)} & \multicolumn{2}{c|}{test clean} & \multicolumn{2}{c}{test other} \\
            \cline{2-7}
            ~&unlabeled&labeled&\multicolumn{1}{c|}{offline} & \multicolumn{1}{c|}{online} & \multicolumn{1}{c|}{offline} & \multicolumn{1}{c}{online} \\
		\hline
		\hline
		  SL~\cite{yao2021wenet} &NA & 100 & 8.3& 9.7&21.9&25.3  \\
            SSL~\cite{fu2022ufo2} & 1k & 100 &5.5&6.9&12.8&{\bf 16.8}  \\
            OTF & NA & 100 &{\bf 5.4}&{\bf 6.5}&{\bf 12.7}&{\bf 16.8}  \\
		\hline
	\end{tabular}
	}
        \vspace{-0.12in}
\end{table}

\begin{table}[t]
	\centering  
		\caption{Performance on small Cantonese dataset in CER ($\%$).}
	\label{table6}  
	\resizebox{0.68\linewidth}{!}{
	\begin{tabular}{l|c|c|c|c}  
		\hline  
            \multirow{2}{*}{Methods} & \multicolumn{2}{c|}{training data (h)} & \multicolumn{2}{c}{test Cantonese} \\
            \cline{2-5}
            ~&unlabeled&labeled & \multicolumn{1}{c|}{offline} & \multicolumn{1}{c}{online} \\
		\hline
		\hline
		  SL~\cite{yao2021wenet} &NA & 50 & 16.6&18.2  \\
            SSL~\cite{fu2022ufo2} & 150k & 50 &14.6&16.7  \\
            OTF & NA & 50 &{\bf 14.2}&{\bf 16.3}  \\
		\hline
	\end{tabular}
	}
        \vspace{-0.22in}
\end{table}

\section{Conclusions}
\label{sec:conclusions}
We proposed a novel weight-based fusion method for SL and SSL ASR models via optimal transport, which improved the recognition performance without increasing the inference cost. Extensive experiments on English and Chinese dataset were conducted to verify the effectiveness of the proposed method. However, models with the same architecture are used to test the performance, the fusion of heterogeneous models will be further investigated in our future work.

\vfill\pagebreak

\bibliographystyle{IEEEtran}
\bibliography{mybib}

\end{document}